\documentclass[12pt,preprint]{aastex}

\shortauthors{Okamoto et al.}
\shorttitle{PROMINENCE FORMATION BY EMERGING HELICAL FLUX ROPE}

\begin{document}

\title{PROMINENCE FORMATION ASSOCIATED WITH AN EMERGING HELICAL FLUX ROPE}
\author{\textsc{
Takenori J. Okamoto,$^{1}$
Saku Tsuneta,$^1$
Bruce W. Lites,$^2$
Masahito Kubo,$^2$
Takaaki Yokoyama,$^3$
Thomas E. Berger,$^4$
Kiyoshi Ichimoto,$^5$
Yukio Katsukawa,$^1$
Shin'ichi Nagata,$^5$
Kazunari Shibata,$^5$
Toshifumi Shimizu,$^6$
Richard A. Shine,$^4$
Yoshinori Suematsu,$^1$
Theodore D. Tarbell,$^4$
and Alan M. Title$^4$
}}
\affil{
$^{1}$National Astronomical Observatory, Mitaka, Tokyo, 181-8588, Japan; joten.okamoto@nao.ac.jp\\
$^{2}$High Altitude Observatory, National Center for Atmospheric Research, P.O. Box 3000, Boulder CO 80307-3000, USA.
\footnote{The National Center for Atmospheric Research is sponsored by the National Science Foundation.}\\
$^{3}$Department of Earth and Planetary Science, School of Science, University of Tokyo, Hongo, Bunkyo, Tokyo, 113-0033, Japan.\\
$^{4}$Lockheed Martin Solar and Astrophysics Laboratory, B/252, 3251 Hanover St., Palo Alto, CA 94304, USA.\\
$^{5}$Kwasan and Hida Observatories, Kyoto University, Yamashina, Kyoto, 607-8471, Japan.\\
$^{6}$ISAS/JAXA, Sagamihara, Kanagawa, 229-8510, Japan.
}

\begin{abstract}

The formation and evolution process and magnetic configuration of solar prominences remain unclear.
In order to study the formation process of prominences, we examine continuous observations of a prominence in NOAA AR~10953 with the Solar Optical Telescope on the \emph{Hinode} satellite.
As reported in our previous Letter, we find a signature suggesting that a helical flux rope emerges from below the photosphere under a pre-existing prominence.
Here we investigate more detailed properties and photospheric indications of the emerging helical flux rope, and discuss their relationship to the formation of the prominence.
Our main conclusions are:
(1) A dark region with absence of strong vertical magnetic fields broadens and then narrows in Ca~\textsc{ii}~H-line filtergrams.
This phenomenon is consistent with the emergence of the helical flux rope as photospheric counterparts.
The size of the flux rope is roughly 30,000 km long and 10,000 km wide.
The width is larger than that of the prominence.
(2) No shear motion or converging flows are detected, but we find diverging flows such as mesogranules along the polarity inversion line.
The presence of mesogranules may be related to the emergence of the helical flux rope.
(3) The emerging helical flux rope reconnects with magnetic fields of the pre-existing prominence to stabilize the prominence for the next several days.
We thus conjecture that prominence coronal magnetic fields emerge in the form of helical flux ropes that contribute to the formation and maintenance of the prominence.

\end{abstract}


\section{Introduction}

Prominences are relatively cool objects embedded in the hotter corona, and are commonly observed above the solar limb in H$\alpha$ emission.
When seen against the solar disk, prominences appear as absorbing dark features  in H$\alpha$ called ``filaments'' since they are elongated and slender (e.g., Lin et al. 2005).
We call them prominences in this paper regardless of the observed disk positions.

Although it is known that prominences are supported by coronal magnetic fields against gravity (see references in Tandberg-Hanssen 1995; Martin 1998),
the structure and formation process of magnetic fields in and around a prominence remains unclear.
Since reliable techniques are not yet available to directly image magnetic fields in the corona, it is difficult to investigate the evolution process of prominence magnetic fields.
However, statistical studies by Leroy (1984) indicate that most prominences have the so-called ``inverse-polarity'' configuration (Kuperus \& Tandberg-Hanssen 1967; Kuperus \& Raadu 1974).

Numerous authors have discussed the formation processes of prominences with the inverse-polarity configuration.
The models may be classified as follows:
the ``flux rope'' model (e.g., Rust \& Kumar 1994; Low \& Hundhausen 1995; Low 1996, 2001; Lites 2005; Zhang \& Low 2005),
the ``sheared-arcade'' model (e.g., Pneuman 1983; van Ballegooijen \& Martens 1989, 1990; Antiochos et al. 1994; DeVore \& Antiochos 2000; 
Martens \& Zwaan 2001; Aulanier et al. 2002; Karpen et al. 2003; Mackay \& van Ballegooijen 2005, 2006),
and the ``quadruple magnetic source'' model (e.g., Malherbe \& Priest 1983; Anzer 1990; D\'emoulin \& Priest 1993; Uchida et al. 1999a, 1999b, Hirose et al. 1999).
Past observations have supported the sheared-arcade model in large quiescent prominences (Gaizauskas et al. 2001; Anderson \& Martin 2005)
and active region prominences (Gaizauskas et al. 1997; Chae et al. 2001; Chae 2003)
although Kubo \& Shimizu (2007) suggest that the sheard-arcade model is inconsistent with their observations of photospheric magnetic fields below the prominences.
Uchida et al. (2003) show structures indicated by the quadruple magnetic source model in an active region prominence.
In contrast, no observational evidence of an emerging helical flux rope has yet been presented 
although some numerical simulations have been carried out (e.g., Magara 2006, 2007; Cheung et al. 2007).

However, a recent study of a series of vector magnetograms obtained by the \emph{Hinode} Solar Optical Telescope 
(SOT, Kosugi et al. 2007; Tsuneta et al. 2008; Suematsu et al. 2008; Ichimoto et al. 2008; Shimizu et al. 2008)
was interpreted as the emergence of a helical flux rope under a prominence in an active region (Okamoto et al. 2008, hereafter Paper I). 
This is the first observational study to detail the evolution of the photospheric magnetic field suggesting 
the emergence of a helical flux rope that may be associated with the formation and maintenance process of an active region prominence.
In this paper, we report the more detailed features of the possible emerging helical flux rope and the relationship between the flux rope and the prominence.

\section{Observation and data analysis}

The \emph{Hinode} satellite observed active region NOAA 10953 from 2007 April 28 to May 9.
We obtained continuous observations of the photosphere and chromosphere with the G-band (4305\AA, band width: 8\AA), Ca \textsc{ii}~H-line (3968\AA, band width: 3\AA), 
and H$\alpha$ (6563\AA, band width: 0.1\AA) filters of the SOT/Filtergraph (FG) from 11:39 UT on April 28 to 17:36 UT on April 30.
The field of view is 108\arcsec$\times$108\arcsec (1024$\times$1024 pixel$^2$) in G-band and Ca \textsc{ii} H and 160\arcsec$\times$160\arcsec (1024$\times$1024 pixel$^2$) in H$\alpha$.
The cadence of the observations is 1 minute, with brief interruptions for synoptic and engineering observations.
In addition, at the beginning of this period, we obtained periodic scans with the SOT/Spectro-Polarimeter (SP) of this active region, including the pre-existing prominence.
The multiple scans were performed in the ``Fast Map'' mode, which
has an integration time of 3.2 s for one slit position and a spatial pixel size of 0.32\arcsec.
The average cadence of the scanning was three hours and the field of view was 160\arcsec$\times$160\arcsec (512$\times$512 pixel$^2$).
The SP simultaneously measures the full Stokes profiles of the Fe I lines at 6301.5 and 6302.5 \AA\ with a sampling of 21.6 m\AA.

After 17:36 UT on April 30, we have no SOT data of the prominence. 
However, we have H$\alpha$ data taken by the Solar Magnetic Activity Research Telescope (SMART; UeNo et al. 2004) at Hida Observatory, Kyoto University.
The SMART observes full-disk H$\alpha$ images every day, weather permitting.
We used the SMART data as complementary H$\alpha$ images.

Vector magnetic fields were derived from the calibrated Stokes profiles on the assumption of a Milne-Eddington atmosphere.
The inversion code used here is MEKSY (Yokoyama et al. 2009), which is based on the PIKAIA (e.g., Charbonneau 1995)
and MELANIE\footnote{http://www.hao.ucar.edu/public/research/cic/melanie.html} codes maintained by the High Altitude Observatory (HAO).
MEKSY was developed to meet the requirement of rapidly processing the large-format SP data.
Higher performance is obtained by removing the redundant procedures from the original codes as much as possible.
The azimuth angle of magnetic fields obtained in any Stokes inversion procedure has an ambiguity of 180$^{\circ}$.
The AZAM utility (Lites et al. 1995) is used to resolve the azimuth ambiguity.
The basic premise of this utility is minimization of spatial discontinuities in the field orientation.

Two types of Dopplergrams were also produced by the inversion.
One is a velocity map of magnetic components derived with all the Stokes profiles, and the other is that of non-magnetic atmosphere derived only with the Stokes I profile.
Here we assume that the large component of the Stokes I profile comes from the non-magnetic atmosphere under condition of low filling factor.
There is no absolute velocity reference for these maps, however
Mart\'{\i}nez Pillet et al. (1997) reported that magnetic component in plage regions is red-shifted with 200 m s$^{-1}$.
Therefore, we set a zero level by assuming that the average velocity in the plage region for which vertical field strength is larger than 1000 G is zero in both types of the Dopplergrams.

\section{Results}

In Paper I, we focused on the evolution of the photospheric region under the active-region prominence only with vector magnetograms derived by SP.
We found following four features:
(1) The abutting opposite-polarity regions on the two sides along the polarity inversion line (PIL) first grew laterally in size and then narrowed.
(2) These abutting regions contained weak vertical, but strong horizontal, magnetic fields.
(3) The orientation of the horizontal magnetic field along the PIL gradually changed with time from a normal-polarity configuration to an inverse-polarity one.
(4) The horizontal magnetic field region was blue-shifted.
We call this region the \textit{weak-field region} in Paper I.
Here we show more detailed features of the weak-field region and evolution of the prominence above the region.

\subsection{Ca \textsc{ii}~H-line movie}

\begin{figure}[bhtp]
\epsscale{1.0}
\plotone{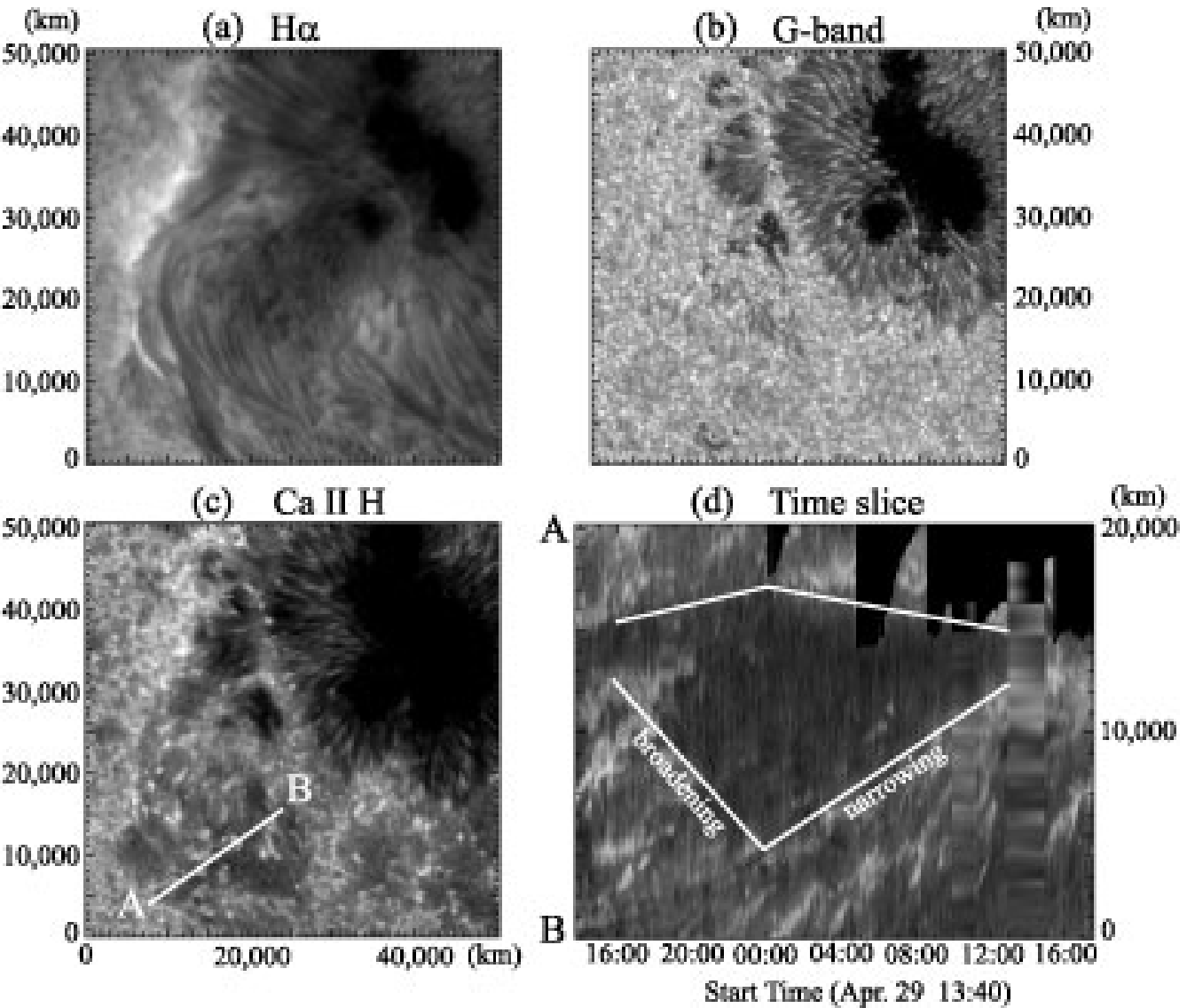}
\caption{
Snapshot images of 3 wavelengths and time-slice image.
{\it a}: H$\alpha$ image.
{\it b}: G-band image.
{\it c}: Ca \textsc{ii}-H image.
{\it d}: Time slice image along a slit A--B in Panel {\it c}.
}
\label{ca_slice}
\end{figure}

We show H$\alpha$, G-band, and Ca \textsc{ii} images in Figure \ref{ca_slice}.
In the Ca \textsc{ii} image, there are numerous bright points around the main sunspot and plage.
These bright points correspond to strong-field magnetic elements,
because the Ca \textsc{ii} bright points are co-spatial with G-band bright points that are manifestation of strong magnetic elements (e.g., Berger \& Title 2001, Ishikawa et al. 2007).
On the other hand, we cannot see the prominence in the Ca \textsc{ii}~H-line image due to the stronger photospheric emission, but there is a filament channel seen as a darker region.
A ``time slice'' is taken along the line (A--B) perpendicular to the prominence in the Ca \textsc{ii} image as shown in Figure \ref{ca_slice}{\it c},
in order to investigate the evolution of the filament channel (Fig. \ref{ca_slice}{\it d}).
The dark region in the Ca \textsc{ii} H time slice clearly broadens and then becomes narrow again over time.
The dark region in Ca \textsc{ii} H (Fig. \ref{ca_slice}{\it c}) has fewer G-band bright points as seen in Figure \ref{ca_slice}{\it b}
indicating the general absence of strong vertical magnetic fields in the filament channel.
Moreover, the width and the time when the width becomes maximum are also consistent with those of the weak-field region obtained by the SP.
Although there are converging motions driven by moat flows from the sunspot toward the PIL outside the weak-field region,
the weak-field region widens, in effect working against the predominant moat flows.

\subsection{Interaction with granules and mesogranules}

We study the physical parameters from the Milne-Eddington inversion of the SP data in the weak-field region. 
To examine the property of the weak-field region, we divided the region near the polarity inversion line into three parts in terms of the vertical magnetic field strength.
The threshold is $\pm 650$ Gauss, and the region with vertical field strength smaller than the threshold is the weak-field region.
Outside the weak-field region, the plage region is located on the east side, and the sunspot is on the west side.
The magnetic properties of the three regions drawn from the SP data scanned around 5:00 UT on April 30 are tabulated in Table \ref{pil}.

\begin{table*}[htbp]
\begin{center}
\begin{tabular}{lcccc}
\\
\hline
    & \multicolumn{2}{c}{Magnetic field strength} & Vertical & Filling \\
    & Vertical & Horizontal & velocity & factor \\
    & (Gauss) & (Gauss) & (m s$^{-1}$) & \\
\hline
East region (plage) & $+1100$ ($\pm 400$) & $400$ ($\pm 200$) & $+200$ ($\pm 500$) & $0.23$ ($\pm 0.20$) \\
weak-field region & \ \ \ \ \ \ $0$ ($\pm 200$) & $650$ ($\pm 150$) & $-300$ ($\pm 200$) & $0.15$ ($\pm 0.05$) \\
West region (sunspot) & $+1000$ ($\pm 400$) & $650$ ($\pm 200$) & $-100$ ($\pm 700$) & $0.30$ ($\pm 0.20$) \\
\hline
\end{tabular}
\caption{
Physical parameters obtained with the Milne-Eddington inversion.
The vertical velocities are those measured from the Stokes Q, U, and V spectra, i.e., the magnetic components.
}
\label{pil}
\end{center}
\end{table*}

\begin{figure}[tbhp]
\epsscale{1.0}
\plotone{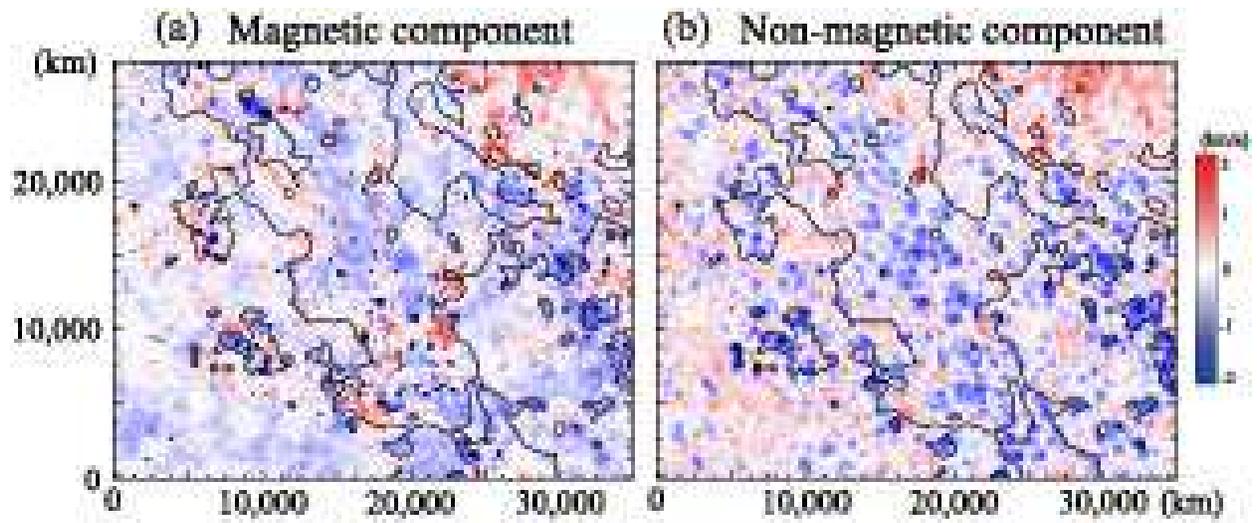}
\caption{
{\it a}: Velocity map of magnetic components derived by the Milne-Eddington inversion.
{\it b}: Velocity map of non-magnetic components. Granular patterns are seen. Field of view same as Panel {\it a}.
}
\label{magnonmag}
\end{figure}

The mean strength of the horizontal magnetic fields in the weak-field region is 650 G.
This value is larger than the strength of the photospheric equipartition field ($\sim$400 G),
for which magnetic energy is comparable to the kinetic energy of the granular convection.
The magnetic field strength is thus strong enough to somewhat suppress the surrounding granular convection.
Normal granules are seen in the weak-field region as seen in the G-band movie.
Figure \ref{magnonmag} shows the Dopplergrams of the magnetic components and the non-magnetic components.
We can see the granular pattern in the map of the non-magnetic velocity, whereas no similar patterns are seen in the map of the magnetic velocity.
That is consistent with the small filling factor in the weak-field region.
The filling factor is defined as the ratio of the area occupied by magnetic fields with the size of the point spread function (Airy disk size).
The actual area of the magnetic component is so small that the magnetic fields in the weak-field region do not suppress the granular convection.

\begin{figure}[bhtp]
\epsscale{1.0}
\plotone{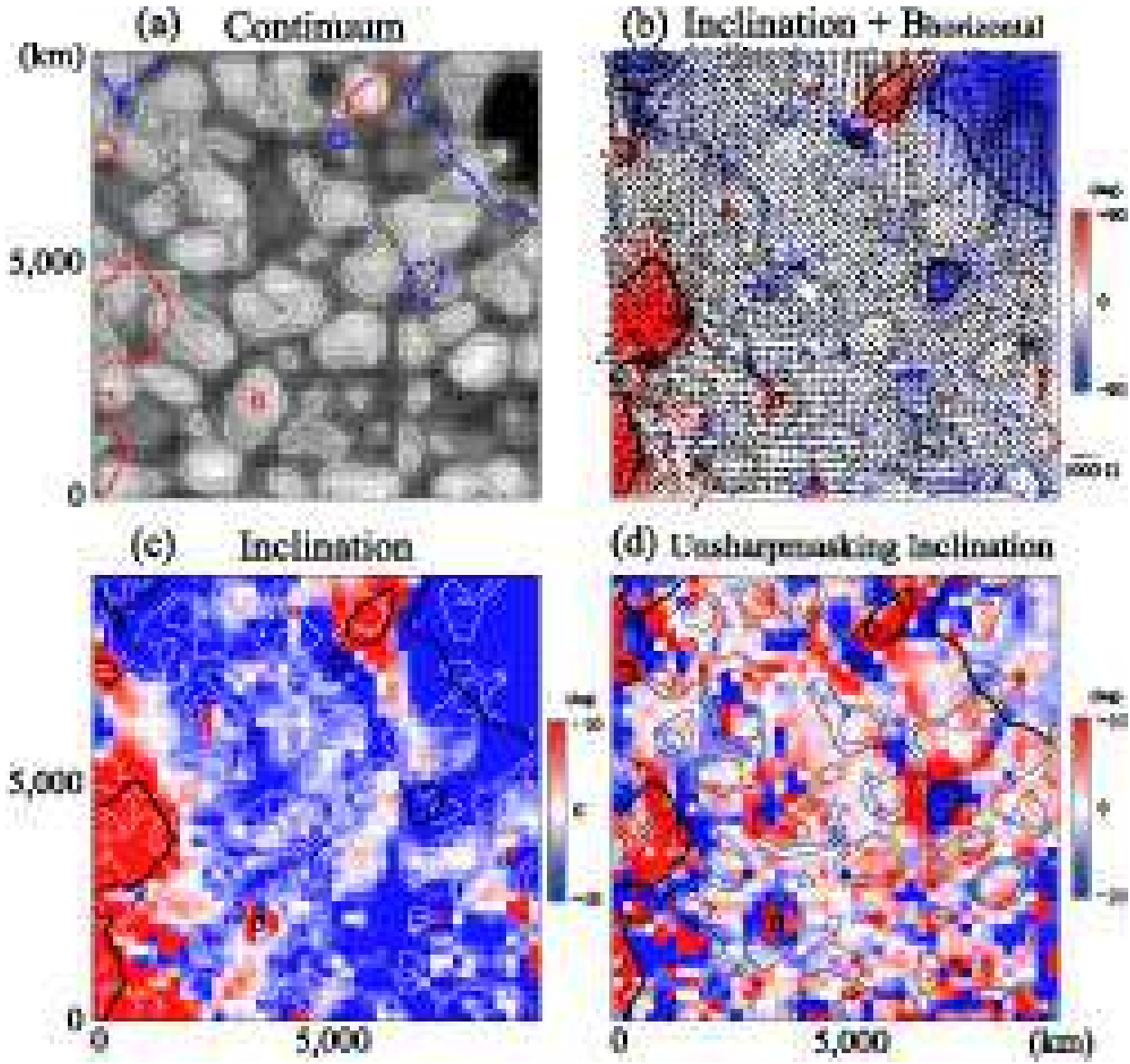}
\caption{
Close-up maps of the weak-field region.
The location of the FOV is (22E, 13S). North is up and east is to the left.
\textit{a}: Continuum intensity map. White lines indicate bright contours of granules.
Red and blue lines show the boundary of $\pm$650-G vertical magnetic field strength, respectively.
\textit{b}: Inclination map with horizontal magnetic fields indicated by arrows.
Red indicates positive polarity (toward us) and blue indicates negative polarity (away from us) in the local frame.
\textit{c}: Inclination map; The color table saturates at $\pm 10^{\circ}$ with respect to the solar surface.
\textit{d}: Unsharp-masking inclination map; The color table saturates
at $\pm 10^{\circ}$.
In Panels \textit{b, c, d}, thin and thick black lines are the same as the white lines and red/blue lines in Panel {\it a}, respectively.
}
\label{gran_hori1}
\end{figure}

\begin{figure}[bhtp]
\epsscale{1.0}
\plotone{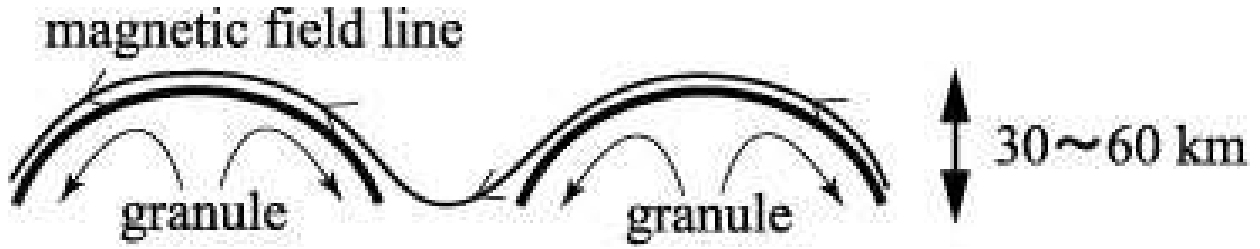}
\caption{
Schematic image of the interaction between granules and horizontal magnetic fields.
}
\label{gran_hori2}
\end{figure}

We take a closer look at the possible interaction between the weak-field region and granular convection.
Figure \ref{gran_hori1} shows a close-up view of the weak-field region.
Figure \ref{gran_hori1}{\it a} is the intensity map. Granular cells are outlined with white contours.
Figure \ref{gran_hori1}{\it b} is the inclination map of the magnetic
fields with the horizontal magnetic fields indicated by arrows.
An inclination of $-90^{\circ}$ ($+90^{\circ}$) represents vertical magnetic fields oriented away from (towards) the local solar surface,
and 0$^{\circ}$ corresponds to a horizontal orientation.
The horizontal magnetic fields are independent of the granular patterns.
The angle of inclination is generally near 0$^{\circ}$ with respect to the solar surface in the weak-field region: we emphasize the deviations from 0$^{\circ}$ in Figure \ref{gran_hori1}{\it c}.
We can see an indication that the inclination in the southwest region of the granules is slightly positive and that in the northeast side is slightly negative.
This reflects the granular shape.
A granule has local maximum height around the center of the granule and minimum height along the boundary with other granules (Fig. \ref{gran_hori2}).
We calculated the height differences across the granules based on the deflection of the magnetic vector.
The height of some granules is 30--60 km, which is in line with numerical simulations of solar granulation by Stein \& Nordlund (1998) and Cheung et al. (2007).
We can determine the horizontal direction of magnetic fields with the relationship between the granular height and the inclination of the magnetic fields.
The direction of the horizontal magnetic fields that is consistent with the granular height is the same as the solution obtained by the AZAM code.

\begin{figure}[bhtp]
\epsscale{1.0}
\plotone{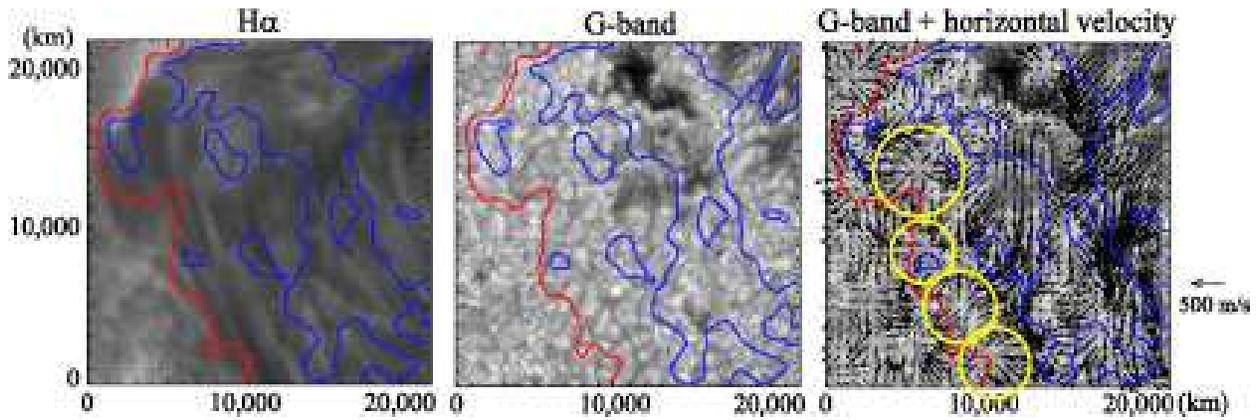}
\caption{
H$\alpha$, G-band, and horizontal flow maps.
Red and blue lines show the boundary of $\pm$650 Gauss of vertical magnetic field strength, respectively.
Yellow circles indicate mesogranules under the prominence.
}
\label{mesogran}
\end{figure}

Furthermore, we examine the relationship of magnetic fields to photospheric motions.
From the time series of the G-band images we calculate the horizontal velocity on the photosphere using Local Correlation Tracking (LCT;
e.g., November \& Simon 1988, Strous et al. 1996, Shine et al. 2000).
The results derived from the LCT method depend on the FWHM apodization (Title et al. 1989, Berger et al. 1998) of the spatial windows used to
correlate successive images in the time series.
We used an apodization of 1.6\arcsec and a 2-hour time average smoothing to measure the flow patterns exclusive of granulation flows.
In Figure \ref{mesogran}, we do not detect any shear motions near the PIL in the horizontal flow map.
Moreover, diverging flows were dominant rather than the converging motions. 
This finding is consistent with the Ca \textsc{ii} H movie (Fig. \ref{ca_slice}).
Since the diverging flows have a roughly circular structure with radii of 3,000--5,000 km, they are consistent with mesogranules (Fig. \ref{mesogran}).
These mesogranules are located along the PIL. 

\subsection{Formation and maintenance of prominence}

\begin{figure}[bhtp]
\epsscale{1.0}
\plotone{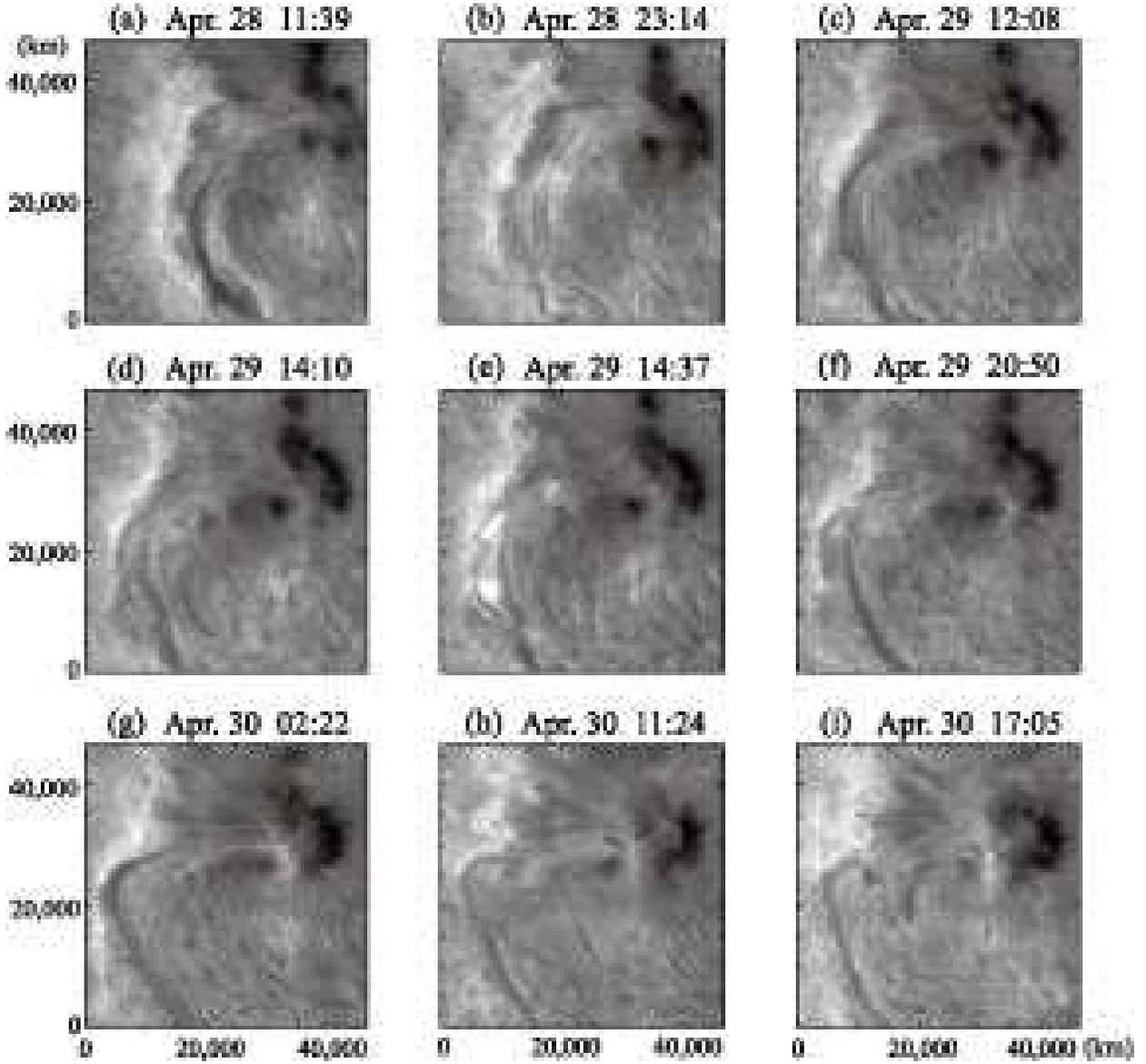}
\caption{
Time series of H$\alpha$ images of the SOT. North is up and east is to the left. Small tickmarks indicate 5,000 km.
}
\label{halphas}
\end{figure}

\begin{figure}[bhtp]
\epsscale{1.0}
\plotone{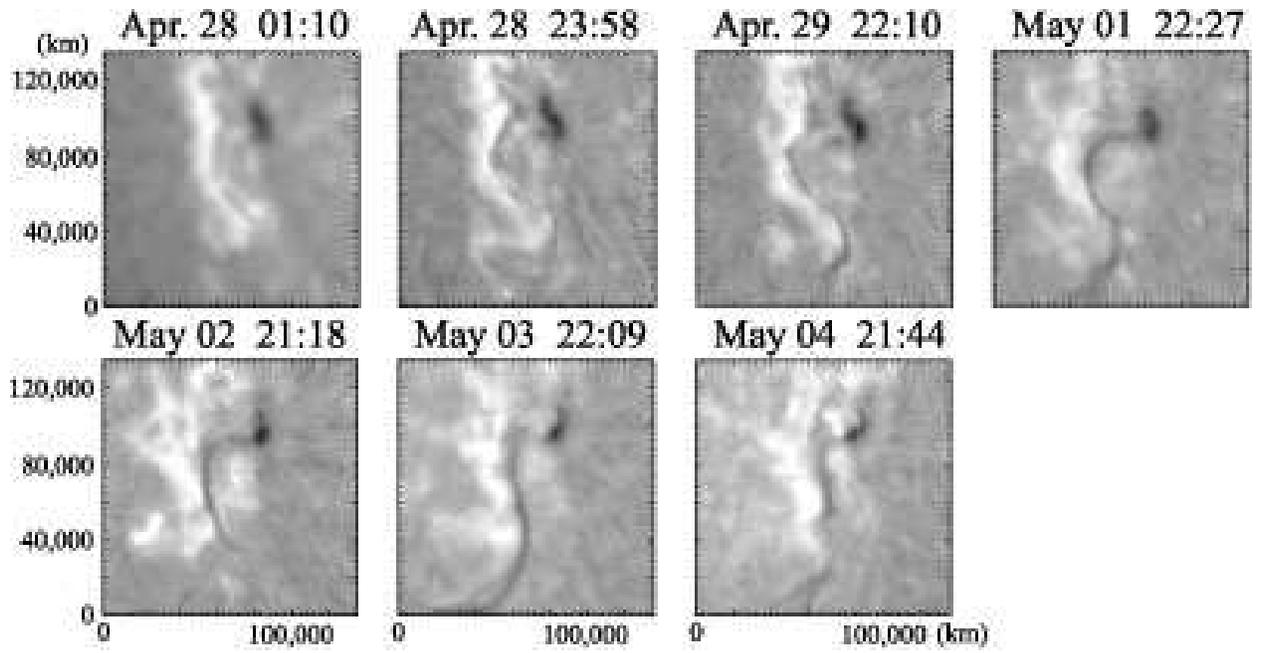}
\caption{
Time series of the active region images in H$\alpha$ obtained with the SMART. North is up and east is to the left. Large tickmarks indicate 20,000 km.
}
\label{smarts}
\end{figure}

The prominence already existed at the beginning of the SOT observation period on April 28 (Fig. \ref{halphas}{\it a}).
In the SOT H$\alpha$ movie we can see that the prominence evolves significantly as described below.
After about 12 hours from the start of the SOT observations, we notice the prominence fragment (Fig. \ref{halphas}{\it b}).
During April~29 the prominence evolves to a continuous relatively straight structure (Fig. \ref{halphas}{\it c}),
but it then loses clear shape (Fig. \ref{halphas}{\it d}).
The prominence becomes a continuous long structure again in Figure \ref{halphas}{\it f}.
Then, it continues to have essentially the same appearance for one day after that (Fig. \ref{halphas}{\it g--i}).
Since we have no SOT H$\alpha$ data after 17:37 UT on April 30, we use the SMART H$\alpha$ data instead.
Although the spatial resolution is lower, we can adequately recognize the prominence shape.
The SMART data show that the prominence keeps the continuous structure for several days after May 1 (Fig. \ref{smarts}).

Based on these observations, we find that there would be a point at which the complex prominence structure becomes simple.
That time is estimated to be around 21:00 UT on April 29 from the examination of Figure \ref{halphas}.

\begin{figure}[bhtp]
\epsscale{1.0}
\plotone{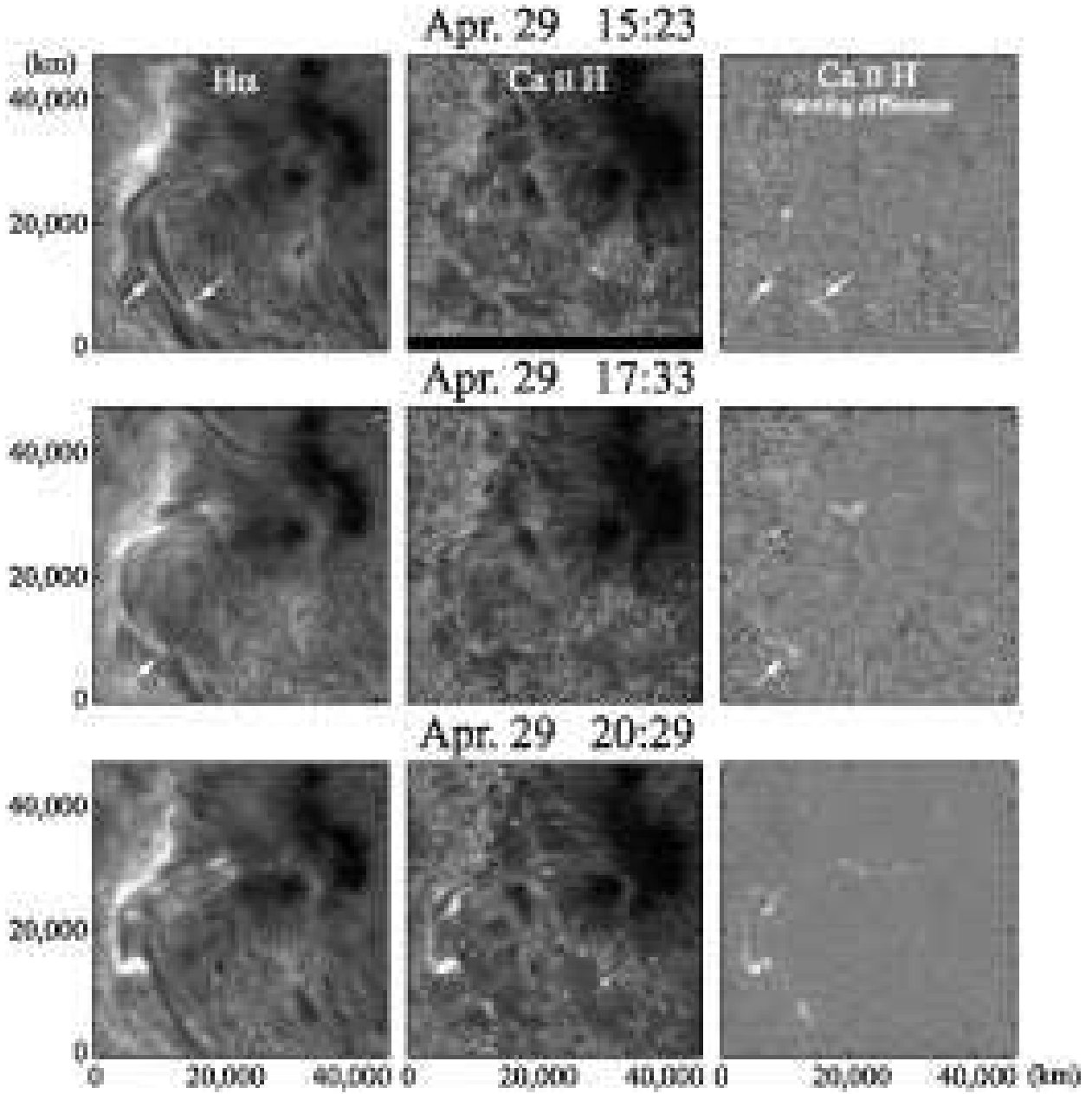}
\caption{
Time series of the active region images in H$\alpha$ and Ca \textsc{ii} H lines.
North is up and east is to the left. Small tickmarks indicate 5,000 km.
Running-difference between consecutive images to emphasize the brightenings and darkenings are shown.
}
\label{haca_brightening}
\end{figure}

According to the Ca \textsc{ii} and H$\alpha$ movies, we can see frequent brightenings along the prominence (Fig. \ref{halphas}{\it e} and Fig. \ref{haca_brightening}).
In particular, brightenings occur around the gap between two previous prominence sections before the change in appearance around 21:00 UT on April 29.
Such transient brightenings in Ca \textsc{ii} appear to be the result of magnetic reconnection between the two or more flux ropes in the corona
forming the complex structure of the prominence seen in Figure \ref{halphas}.
The brightenings are not seen after 21:00 UT on April 29, when we begin to see a simple structure.
The replacement of new simple flux rope structure from the old complex structure may explain the disappearance of such transient heating due to magnetic reconnection.

\section{Discussion}

In Paper I, we postulated that the weak-field region was a signature of a helical flux rope emerging from below the photosphere.
In this section, we discuss the property of the possible emerging helical flux rope with the observational results,
and also the relationship between the emerging helical flux rope and evolution of the prominence.

\subsection{Emergence of a helical flux rope}

\begin{figure}[bhtp]
\epsscale{1.0}
\plotone{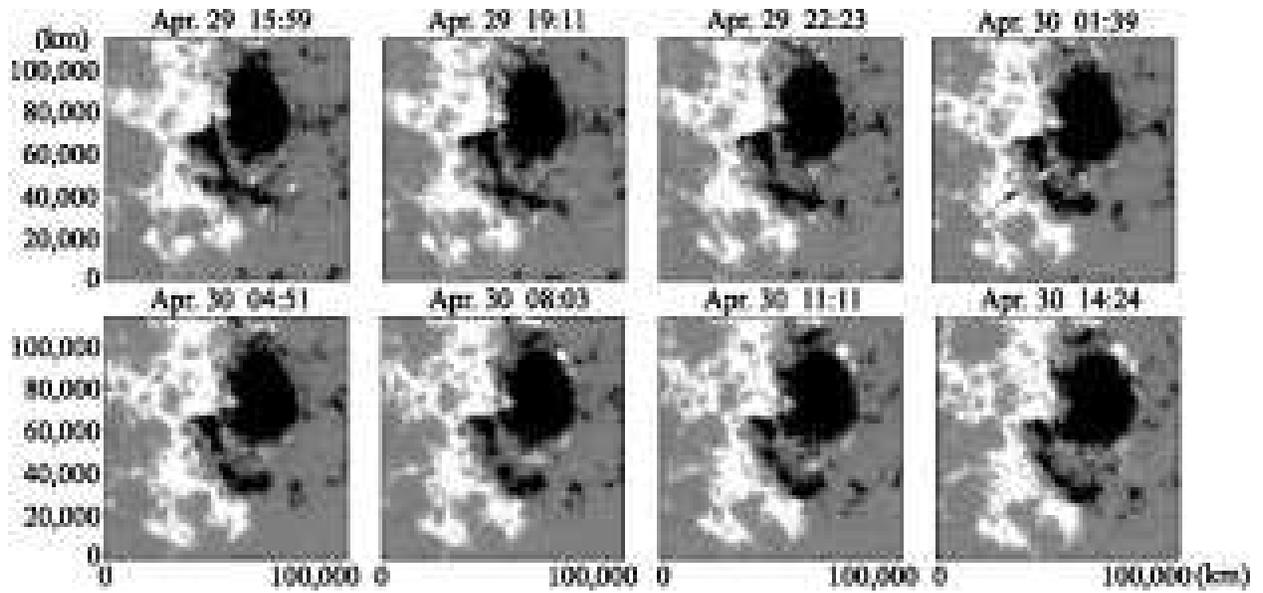}
\caption{
Line-of-sight magnetograms in the active region taken with the
Michelson Doppler Imager (MDI) on board \emph{SOHO}.
White and black colors indicate magnetic flux.
The regions with more than 200 G and less than $-200$ G are saturated.
}
\label{mdi}
\end{figure}

As described in Paper I, the weak-field region is occupied by weak vertical, but strong horizontal, magnetic fields,
and the region first grows laterally in size and then narrows observed in time-series of vector magnetograms derived by SP.
The strength of the magnetic fields is about 650 Gauss, which is weaker than that outside of the weak-field region.
Moreover, we show some features in the filtergraph observations:
(1) A dark region with absence of strong vertical fields broadens and narrows in the movie of Ca \textsc{ii}.
(2) Diverging flow patterns line up along the weak-field region.
These features are consistent with a hypothesis of a horizontally-dominant flux rope emerging from below the photosphere.
However, using only the filtergraph observations, we cannot determine the existence of a flux emergence with confidence.
Observations comprising only line-of-sight magnetograms suffer from the same limitation.
Now we consider the possibility of detection of an emerging helical flux rope without vector magnetograms.
We have the full-disk magnetograms of the line-of-sight component taken with the \emph{SOHO}/MDI (Domingo et al. 1995; Scherrer et al. 1995), which are frequently used
for evolutionary studies of the magnetic field.
The cadence is 96 minutes.
Figure \ref{mdi} is the evolution of the magnetic fields in the active region including the main sunspot and the prominence reported here.
The white (black) indicates positive (negative) polarity.
We see that a gray region (weak field corridor in a study of Klimchuk 1987) pointed out by the black and white arrows widens and then narrows just as presented above in the SOT data.
The gray region looks like absence of magnetic fields, but the region is actually occupied by horizontal magnetic fields observed with the SP.
Moreover, the orientation of the horizontal magnetic field in the weak-field region gradually changed with time from a normal-polarity configuration to an inverse-polarity one,
thus strongly suggesting the existence of an emerging helical flux rope, although there may be alternative interpretations that we have not considered.
This indicates that observations of vector magnetic fields and Doppler motion are necessary to detect a possible flux rope emergence.

\subsection{Prominence formation}

\begin{figure}[bhtp]
\epsscale{1.0}
\plotone{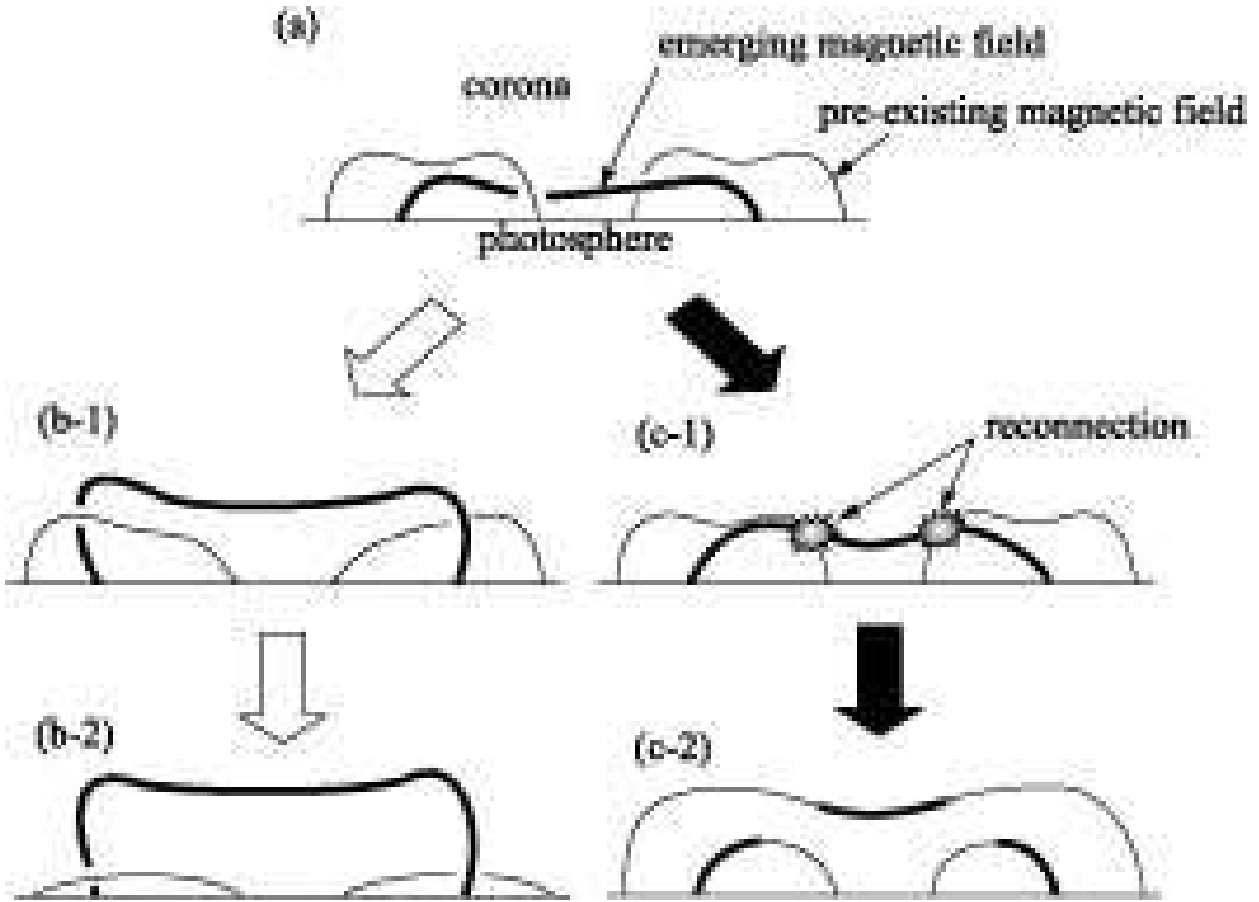}
\caption{
Schematic illustrations of prominence formation from a side view of the PIL.
Thin and thick curved lines indicate magnetic fields of pre-existing prominences and the emerging helical rope, respectively.
Prominence material is not shown and helical configuration of magnetic fields is simplified.
{\it a}: Emergence of the helical flux rope.
{\it Case b}: The emerging flux rope occupies the volume where the pre-existing prominences are located.
{\it Case c}: The emerging flux rope reconnects with magnetic fields of the pre-existing prominences.
}
\label{magform}
\end{figure}

Next, we discuss formation of the prominence, considering the hypothesis of the emerging helical magnetic flux rope.
In our observations, we have a pre-existing, but fuzzy prominence along the PIL.
The prominence develops a more coherent structure in coincidence with the apparent emergence of the flux rope starting at about 16:00 UT on April 29.
Thus, we propose two possibilities to explain this episode in the context of an emerging flux rope.
(1) The emerging helical rope occupies the position of the previous prominence, and provides with the fresh magnetic fields in the corona (Fig. \ref{magform}{\it b}).
As mentioned above, the flux rope could already be partially emerged.
As the emergence continues, the field structure distorts and some of the prominence material drains out 
causing the disappearance of the coherent filament structure, soon replaced by a filament in a slightly different location.
(2) There are remnants of the previous prominence (Fig. \ref{halphas}{\it d}).
If reconnection takes place between the magnetic fields of the remnants and those of the emerging helical rope,
a continuous magnetic field would be formed (Fig. \ref{magform}{\it c}).

Here we mention the possibility the latter process since there are several brightenings along and around the prominence just before this change of the prominence.
We suggest that these brightenings indicate reconnection between the fragmented prominence sections and the emerging flux rope that constructs the longer coherent prominence.
Chae et al. (2001), Chae (2003), and Schmieder (2004) also report that similar brightenings occurred before prominence formation in active regions.
In their cases, two prominences were located in proximity and connected after the brightening events.
They observed magnetic cancellation with sheared converging motion toward the PIL.
The numerical simulations of Aulanier et al. (2006) suggest the possibility that shear motion and reconnection leads to merging of prominences.
However, our data show neither shear nor converging motion under the prominence sections. 
Rather we find diverging motion in the form of mesogranules before the prominence formation.
In our case, we suggest that the emergence of a magnetic flux rope is the dominant process for the formation of the more coherent prominence,
and that granular and mesogranular flows also play a important role in the formation process of prominence.

\subsection{Mass supply}

We point out open issues unanswered in this study.
The first is the mass supply of the prominence.
In this case shown in Figure \ref{magform}{\it b}, the prominence mass might be supplied from below with the emerging helical rope.
On the other hand, in the case of Figure \ref{magform}{\it c},
we have a possibility that the pre-existing prominence provides new magnetic fields with the mass.
The helical rope emerges into the corona, and may reconnect with the magnetic fields of the pre-existing prominence.
If the mass of the pre-existing prominence is not lost during the reconnection process, the mass may be shared in the newly formed prominence.
The emerging helical flux may supply only magnetic fields without mass from below due to the Parker instability (Parker 1955, 1966, Zwaan 1985),
However, the horizontal nature of the emerging flux rope may make the classical Parker instability difficult.
This problem can be explained with the following scenario:
Before the emergence, the flux rope may be distorted by convective motions due to their sub-equipartition field strength.
U-shape structures are constructed along the rope, and the mass in the rope moves to the bottom of the U-shape dips (van Driel-Gesztelyi et al. 2000).
The U-shape dips with the mass are separated from the main body of the flux rope by magnetic reconnection (Spruit et al. 1987).
Therefore, the flux rope has smaller amount of mass, and it can emerge from below the photosphere as shown on this observation.
In this case, the Parker instability is not required during the emergence, and then it could horizontally rise up.
Figure \ref{gran_hori1} shows that the horizontal flux rope is modulated by the granular motion.
This may be related to the process described here.
Since the density in the flux tube decreases, the flux tube may radially shrink.
Hence, the magnetic strength increases, while the filling factor decreases, as our observations seem to indicate.

\subsection{Barbs}

\begin{figure}[bhtp]
\epsscale{1.0}
\plotone{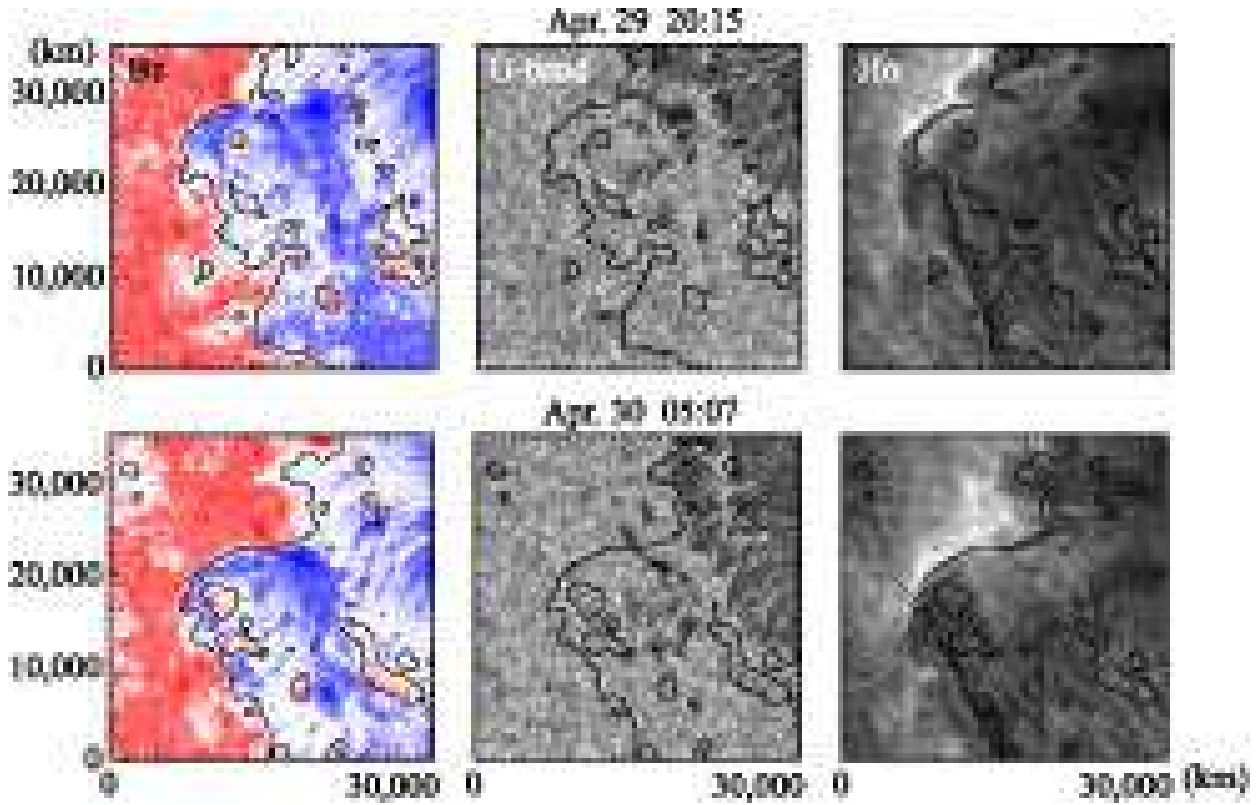}
\caption{
{\it left column}: Magnetograms inferred from SP observations.
Color contour indicates the vertical magnetic strength.
{\it middle column}: G-band images.
{\it right column}: H$\alpha$ images.
The black line in each panel indicates the PIL.
The longest PIL separates the two regions with opposite polarities.
Smaller regions with black circles have minority polarity.
Some endpoints of the prominence are located at such smaller regions.
}
\label{barb}
\end{figure}

The second issue is formation of barbs.
As proposed by Martin \& Echols (1994), barbs are supposedly connected to patches of minority polarity on each side of a prominence.
Recent numerical simulations performed by Magara (2007) showed that the formation of barbs is associated with emerging twisted flux.
The barbs in their simulations had magnetic dips and the configurations were consistent with previous models (Aulanier \& D\'emoulin 1998, van Ballegooijen 2004, L\'opez Ariste et al. 2006).
In our observations, the endpoints of some of the prominence fragments seem to be located at minority polarity sites (Fig. \ref{barb}).
However, it is difficult to accurately locate the endpoints of barbs without H$\alpha$ Dopplergram observations because the upper prominence material obscures the underlying structures.
We do not know the relationship between these potential barb sites and the emergence of the helical rope since the barbs existed before the emergence.

\subsection{Implication of flux rope emergence to dynamo process}

\begin{figure}[bhtp]
\epsscale{1.0}
\plotone{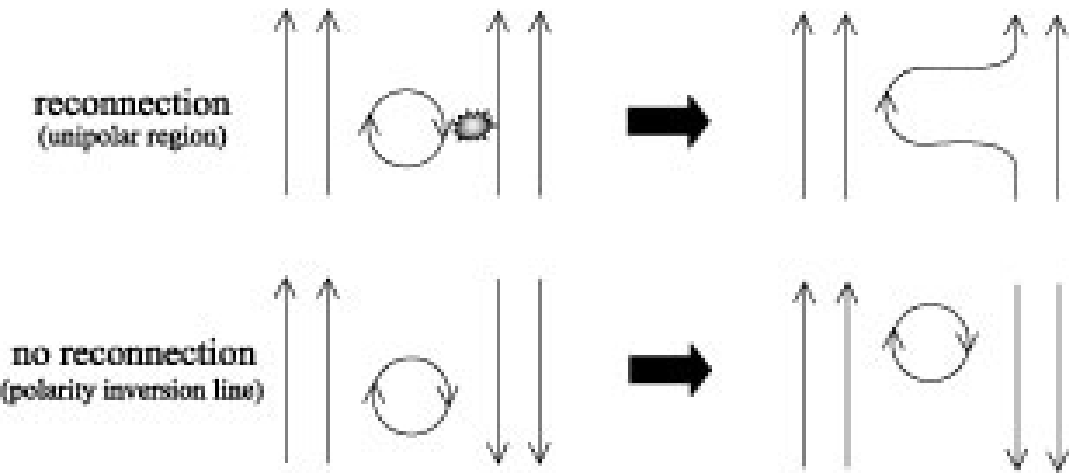}
\caption{
Schematic illustration of the relationship between emerging helical flux rope and vertical magnetic fields.
{\it top}: Case of unipolar vertical magnetic fields. Reconnection occurs between emerging helical flux rope and vertical magnetic fields, and the helical rope is destroyed.
{\it bottom}: Case of PIL. Reconnection does not occur, and the helical rope survives and can emerge.
}
\label{reconnection}
\end{figure}

The emerging helical flux is located along the PIL.
Why is it located at the PIL? Why is it so parallel to the PIL?
If such a helical flux rises up elsewhere, it would reconnect with surrounding unipolar vertical magnetic fields, and become part of vertical magnetic fields (upper panels in Fig. \ref{reconnection}).
On the other hand, if the helical flux rope emerges around the PIL, it will maintain its flux-rope structure (lower panels in Fig. \ref{reconnection}).
There may be a mechanism that helical fluxes are created below the PIL or the helical flux rope emerges everywhere in the active region, 
and most of them except for those appearing along PIL are destroyed by magnetic reconnection with the surrounding field lines.

If emerging helical ropes contribute a great deal to formation of prominences, occurrence of prominences should be modulated by the solar cycle, as are sunspots.
Quiescent prominences have properties considerably different from active-region prominences in recent \emph{Hinode} observations reported by Berger et al. (2008) and Okamoto et al. (2007),
although Martin et al. (2008) suggest that active-region and quiescent prominences have a continuous spectrum of properties.
The shear motion due to the differential rotation may play a major role in the formation of quiescent prominences (Gaizauskas et al. 2001; Anderson \& Martin 2005),
while Hansen \& Hansen (1975) reported that large quiescent prominences, or polar crown prominences, are seen more frequently at solar maximum than at minimum.
Quiet-Sun prominences apparently have a different formation mechanism.
This information may be relevant to the dynamo process.
We should have more observations of both active region and quiescent prominences.

We interpret one set of \emph{Hinode} observations as containing information that is consistent with a flux rope model.
This episode is the only \emph{Hinode} observation made so far for an active region prominence.
We should have more observations to confirm whether this episode is a common or unique phenomenon.
Another episode of emergence observed by \emph{Hinode} has been described briefly by Lites (2008), and will be discussed further in a following paper.
That observation generally supports the conclusions derived from this work.

\

The authors thank the referee for many useful comments.
\emph{Hinode} is a Japanese mission developed and launched by ISAS/JAXA, with NAOJ as domestic partner and NASA and STFC (UK) as international partners.
It is operated by these agencies in co-operation with ESA and NSC (Norway).
This work was supported by KAKENHI (20068097) and carried out at the NAOJ Hinode science center, which was supported by the Grant-in-Aid for Creative Scientific Research ``The Basic Study of
Space Weather Prediction'' from MEXT, Japan (Head Investigator: K. Shibata), generous donation from the Sun Microsystems Inc., and NAOJ internal funding.
The FPP project at LMSAL and HAO is supported by NASA contract NNM07AA01C.

\begin{thebibliography}

\bibitem[---]{---}
Anderson, M., \& Martin, S. F. 2005, in Large-scale Structures and their Role in Solar Activity, eds. K. Sankarasubramanian, M. Penn \& A. Pevtsov, ASP Conf. Ser., 346, 201

\bibitem[Antiochos, Dahlburg, \& Klimchuk 1994]{ant94}
Antiochos, S. K., Dahlburg, R. B., \& Klimchuk, J. A. 1994, ApJ, 420, L41

\bibitem[---]{---}
Anzer, U. 1990, Sol. Phys., 130, 403

\bibitem[---]{---}
Aulanier, G., \& D\'emoulin, P. 1998, A\&A, 329, 1125

\bibitem[Aulanier et al. 2002]{aul02}
Aulanier, G., DeVore, C. R., \& Antiochos, S. K. 2002, 567, L97

\bibitem[---]{---}
------. 2006, ApJ, 646, 1349

\bibitem[---]{---}
Berger, T. E., L\"{o}fdahl, M. G., Shine, R. A., \& Title, A. M. 1998, ApJ, 495, 973

\bibitem[---]{---}
Berger, T. E., \& Title, A. M. 2001, ApJ, 553, 449

\bibitem[---]{---}
Berger, T. E., Shine, R. A., Slater, G. L., Tarbell, T. D., Title, A. M., Okamoto, T. J., Ichimoto, K., Katsukawa, Y., Suematsu, Y., Tsuneta, S., 
Lites, B. W., \& Shimizu, T. 2008, ApJ, 676, L89

\bibitem[---]{---}
Chae, J., Wang, H., Qiu, J., Goode, P. R., Strous, L., \& Yun, H. S. 2001, ApJ, 560, 476

\bibitem[---]{---}
Chae, J. 2003, ApJ, 584, 1084

\bibitem[---]{---}
Charbonneau, P. 1995, ApJS, 101, 309

\bibitem[Cheung et al. 2007]{che07}
Cheung, M. C. M., Sch\"{u}ssler, M., \& Moreno-Insertis, F. 2007, A\& A, 467, 703



\bibitem[---]{---}
D\'emoulin, P., \& Priest, E. R. 1993 Sol. Phys., 144, 283


\bibitem[DeVore and Antiochos 2000]{dev00}
DeVore, C. R., \& Antiochos, S. K. 2000, ApJ, 539, 954

\bibitem[---]{---}
Domingo, V., Fleck, B., \& Poland, A. I. 1995, Sol. Phys., 162, 1

\bibitem[Gaizauskas et al. 1997]{gai97}
Gaizauskas, V., Zirker, J. B., Sweetland, C., \& Kovacs, A. 1997, ApJ, 479, 448

\bibitem[---]{---}
Gaizauskas, V., Mackay, D. H., \& Harvey, K. L. 2001, ApJ, 558, 888

\bibitem[---]{---}
Hansen, R., \& Hansen, S. 1975, Sol. Phys., 44, 225


\bibitem[---]{---}
Hirose, S., Uchida, Y., Uemura, S., Yamaguchi, T., \& Cable, S. 1999, ApJ, 551, 586

\bibitem[Ichimoto et al. 2008]{ich08}
Ichimoto, K., Lites, B., Elmore, D., Suematsu, Y., Tsuneta, S., et al. 
2008, Sol. Phys., 249, 233


\bibitem[---]{---}
Ishikawa, R., Tsuneta, S., Kitakoshi, Y., Katsukawa, Y., Bonet, J. A., Vargas Dom\'{\i}nguez, S., Rouppe van der Voort, L. H. M., Sakamoto, Y., \& Ebisuzaki, T. 2007, A\&A, 472, 911


\bibitem[Karpen et al. (2003)]{kar03}
Karpen, J. T., Antiochos, S. K., Klimchuk, J. A., \& MacNeice, P. J. 2003, ApJ, 593, 1187


\bibitem[---]{---}
Klimchuk, J. A. 1987, ApJ, 323, 368

\bibitem[Kosugi et al. 2007]{kos07}
Kosugi, T., Matsuzaki, K., Sakao, T., Shimizu, T., Sone, Y., et al. 2007, Sol. Phys., 243, 3

\bibitem[Kubo and Shimizu 2007]{kub07}
Kubo, M., \& Shimizu, T. 2007, ApJ, 671, 990

\bibitem[Kuperus \& Raadu 1974]{kup74}
Kuperus, M., \& Raadu, M. A. 1974, A\&A, 31, 189

\bibitem[Kuperus \& Tandberg-Hanssen 1967]{kup67}
Kuperus, M., \& Tandberg-Hanssen, E. 1967, Sol. Phys., 2, 39


\bibitem[Leroy et al. 1984]{ler84}
Leroy, J. L., Bommier, V., \& Sahal-Br\'{e}chot, S. 1984, A\& A, 131, 33

\bibitem[---]{---}
Lin, Y., Engvold, O., Rouppe van der Voort, L., Wiik, J. E., \& Berger, T. E. 2005, Sol. Phys., 226, 239


\bibitem[Lites 1995]{lit95}
Lites, B. W., Low, B. C., Mart\'{\i}nez Pillet, V., Seagraves, P., Skumanich, A., Frank, Z., Shine, R. A., \& Tsuneta, S. 1995, ApJ, 446, 877

\bibitem[Lites 2005]{lit05}
Lites, B. W. 2005, ApJ, 622, 1275

\bibitem[---]{---}
Lites, B. W. 2008, Space Sci. Rev. (DOI 10.1007/s11214-008-9437-x)

\bibitem[---]{---}
L\'opez Ariste, A., Aulanier, G., Schmieder, B., \& Sainz Dalda, A. 2006, A\&A, 456, 725

\bibitem[Low 1996]{low96}
Low, B. C. 1996, Sol. Phys., 167, 217

\bibitem[Low 2001]{low01}
------. 2001, J. Geophys. Res., 106, 25141

\bibitem[Low \& Hundhausen 1995]{low95}
Low, B. C., \& Hundhausen, J. R. 1995, ApJ, 443, 818

\bibitem[Mackay and van Ballegooijen 2005]{mac05}
Mackay, D. H., \& van Ballegooijen, A. A. 2005, ApJ, 621, L77

\bibitem[Mackay and van Ballegooijen 2006]{mac06}
------. 2006, ApJ, 641, 577

\bibitem[---]{---}
Magara, T. 2006, ApJ, 653, 1499

\bibitem[---]{---}
------. 2007, PASJ, 59, L51

\bibitem[---]{---}
Malherbe, J. M., \& Priest, E. R. 1983, A\&A, 123, 80

\bibitem[Martens \& Zwaan (2001)]{mar01}
Martens, P. C., \& Zwaan, C. 2001, ApJ, 558, 872

\bibitem[---]{---}
Martin, S. F. \& Echols, C. R. 1994, in Kluwer Acad. Publ., Dordrecht, Holland, eds. R. J. Rutten \& C. J. Schrijver, 339

\bibitem[Martin 1998]{mar98}
Martin, S. F. 1998, Sol. Phys., 182, 107

\bibitem[---]{---}
Martin, S. F., Lin, Y., \& Engvold, O. 2008, Sol. Phys., 250, 31

\bibitem[---]{---}
Mart\'{\i}nez Pillet, V., Lites, B. W., \& Skumanich, A. 1997, ApJ, 474, 810

\bibitem[---]{---}
November, L. J., \& Simon, G. W. 1988, ApJ, 333, 427

\bibitem[Okamoto et al. 2007]{oka07}
Okamoto, T. J., Tsuneta, S., Berger, T. E., Ichimoto, K., Katsukawa, Y., Lites, B. W., Nagata, S., Shibata, K., Shimizu, T., Shine, R. A., Suematsu, Y., Tarbell, T. D., \& Title, A. M.
2007, Science, 318, 1577

\bibitem[---]{---}
Okamoto, T. J., Tsuneta, S., Lites, B. W., Kubo, M., Yokoyama, T., Berger, T. E., Ichimoto, K., Katsukawa, Y., Nagata, S., Shibata, K., Shimizu, T., Shine, R. A., Suematsu, Y., Tarbell, T. D., \& Title, A. M.
2008, ApJ, 673, L215 (Paper I)

\bibitem[---]{---}
Parker, E. N. 1955, ApJ, 121, 491

\bibitem[---]{---}
------. 1966, ApJ, 145, 811

\bibitem[Pneuman 1983]{pne83}
Pneuman, G. W. 1983, Sol. Phys., 88, 219




\bibitem[Rust and Kumar 1994]{rus94}
Rust, D. M., \& Kumar, A. 1994, Sol. Phys., 155, 69

\bibitem[---]{---}
Scherrer, P. H., Bogart, R. S., Bush, R. I., Hoeksema, J. T., Kosovichev, A. G., Schou, J., et al. 1995, Sol. Phys., 162, 129

\bibitem[---]{---}
Schmieder, B., Mein, N., Deng, Y., Dumitrache, C., Malherbe, J. M., Staiger, J., \& DeLuca, E. E. 2004, Sol. Phys., 223, 119

\bibitem[Shimizu et al. 2007]{shi07}
Shimizu, T., Nagata, S., Tsuneta, S., Tarbell, T., Edwards, C., et al.
2008, Sol. Phys., 249, 221

\bibitem[---]{---}
Shine, R. A., Simon, G. W., \& Hurlburt, N. E. 2000, Sol. Phys., 193, 313


\bibitem[---]{---}
Spruit, H. C., Title, A. M., \& van Ballegooijen, A. A. 1987, Sol. Phys., 110, 115

\bibitem[---]{---}
Stein, R. F., \& Nordlund, \AA. 1998, ApJ, 499, 914

\bibitem[---]{---}
Strous, L. H., Scharmer, G., Tarbell, T. D., Title, A. M., \& Zwaan, C. 1996, A\&A, 306, 947

\bibitem[Suematsu et al. 2007]{sue07}
Suematsu, Y., Tsuneta, S., Ichimoto, K., Shimizu, T., Otsubo, et al. 
2008, Sol. Phys., 249, 197



\bibitem[Tandberg-Hanssen 1995]{tan95}
Tandberg-Hanssen, E. 1995, The Nature of Solar Prominences (Dordrecht: Kluwer)

\bibitem[---]{---}
Title, A. M., Tarbell, T. D., Topka, K. P., Ferguson, S. H., Shine, R. A., \& the SOUP Team 1989, ApJ, 336, 475

\bibitem[Tsuneta et al. 2007]{tsu07}
Tsuneta, S., Suematsu, Y., Ichimoto, K., Shimizu, T., Otsubo, M., et al.
2008, Sol. Phys., 249, 167

\bibitem[---]{---}
Uchida, Y., Fujisaki, K., Morita, S., Torii, M., Hirose, S., \& Cable, S. 1999a, PASJ, 51, 53

\bibitem[---]{---}
Uchida, Y., Hirose, S., Cable, S., Morita, S., Torii, M., Uemura, S., \& Yamaguchi, T. 1999b, PASJ, 51, 553

\bibitem[---]{---}
Uchida, Y., Title, A., Kubo, M., Tanaka, T., Morita, S., \& Hirose, S. 2003, PASJ, 55, 305

\bibitem[---]{---}
UeNo, S., Nagata, S., Kitai, R., \& Kurokawa, H. 2004, in ASP Conf. Ser. 325, The Solar-B Mission and the Forefront of Solar Physics, eds. T. Sakurai \& T. Sekii (San Francisco: ASP), 319


\bibitem[van Ballegooijen and Martens 1989]{vba89}
van Ballegooijen, A. A., \& Martens, P. C. H. 1989, ApJ, 343, 971

\bibitem[van Ballegooijen and Martens 1990]{vba90}
------. 1990, ApJ, 361, 283

\bibitem[---]{---}
van Ballegooijen, A. A. 2004, ApJ, 612, 519

\bibitem[---]{---}
van Driel-Gesztelyi, L., Malherbe, J.-M., \& D\'emoulin, P. 2000 A\&A, 364, 845


\bibitem[---]{---}
Yokoyama, T., Katsukawa, Y., \& Shimojo, M. 2009, in preparation

\bibitem[Zhang and Low 2005]{Zha05}
Zhang, M., \& Low, B. C. 2005, ARAA, 43, 103

\bibitem[]{}
Zwaan, C. 1985, Sol. Phys., 100, 397




\end{thebibliography}
\end{document}